\documentclass[11pt]{elsart}
\usepackage[dvips]{graphicx}

\begin{document}
\begin{frontmatter}

\title{Complex Network Properties of Chinese Natural Science Basic Research}

\author{Jianguo Liu,\corauthref{jg}}
\corauth[jg]{Corresponding author. Tel. (+86)13050537943.}
\ead{liujg004@yahoo.com.cn}
\author{Yanzhong Dang and}
%\ead{yzhdang@dlut.edu.cn}
\author{Zhongtuo Wang}
%\ead{wangzt@dlut.edu.cn}

\address{
Institute of System Engineering, Dalian University of Technology\\
2 Ling Gong Rd., Dalian 116023, Liaoning, P. R. China}%
%

%\author{Tao Zhou}
%\address{
%Department of Modern Physics, University of Science and Technology
%of China, Hefei Anhui, 230026, P. R. China}

\begin{abstract}
In this paper, we studied the research areas of Chinese natural
science basic research from a point view of complex network. Two
research areas are considered to be connected if they appear in
one fund proposal. The explicit network of such connections using
data from 1999 to 2004 is constructed. The analysis of the real
data shows that the degree distribution of the {\bf research areas
network} (RAN) may be better fitted by the exponential
distribution. It displays small world effect in which randomly
chosen pairs of research areas are typically separated by only a
short path of intermediate research areas. The average distance of
RAN decreases with time, while the average clustering coefficient
increases with time, which indicates that the scientific study
would like to be integrated together in terms of the studied
areas. The relationship between the clustering coefficient $C(k)$
and the degree $k$ indicates that there is no hierarchical
organization in RAN.
\end{abstract}
%\pacs{89.20.Hh, 89.75.Hc, 89.75.Da}
%89.20.Hh World Wide Web, Internet
%89.75.Da Systems obeying scaling laws
%89.75.Fb Structures and organization in complex systems
%89.75.-k Complex systems
%89.75.Hc Networks and genealogical trees

\date{}
\end{frontmatter}

%%%%%%%%%%%%%%%%%%%%%%%%%%%%%%%%%%%%%%%%%%%%%%%%%%%%%%%%%%%%%%%%%
%%%%%%%%%%%%%%%%%%%%%%%%%%%%%%%%%%%%%%%%%%%%%%%%%%%%%%%%%%%%%%%%%
%\vskip -0.5cm\color{Blue}
%\vbox to 0pt{\kern -14cm {
%\noindent \small \copyright 2005
%{\em Elsevier Science B.V. All rights reserved}\\
%{\em Physica A}, submitted.}
%\vss}\color{Black}

%%%%%%%%%%%%%%%%%%%%%%%%%%%%%%%%%%%%%%%%%%%%%%%%%%%%%%%%%%%%%%%%%%%%
\section{Introduction}
In the past few years there has been a growing interest in the
study of complex networks. The boom has two reasons: the existence
of interesting applications in several biological, sociological,
technological and communications systems and the availability of a
large amount of real data
\cite{WS98,DMbook,DMW2003,OC2004,DRH2004,XCHH2004,AB02,DM02,New,XFWang2002,XFWangC2003}.
Recent works on the mathematics of networks have been driven
largely by the observed properties of actual networks and the
studies on network
dynamics\cite{D1,D2,D3,D4,D5,D6,D7,D8,D9,D10,D11},
optimization\cite{O1,O2,O3,O4,O5,O6}, and
evolutionary\cite{Strogatz,New00,XL2003,Kast,DM00,Klern,MMP,M99,ZWJX,ZYW,zzz,bw,AHAS,ustc3,CF,CS,DGM,RB,ustc1,ustc2}.
It also makes sense to examine simultaneously data from different
kinds of networks. Recent approaches with methodology rooted in
statistical physics focus on large networks, searching for
universality both in the topology of the real networks and in the
dynamics governing their evolution \cite{BNRT2002}.
%\footnote{search the published paper entitle "evolution of the social network of scientific collaborations"}.
These combined theoretical and empirical results have opened
unsuspected directions for researches and a wealth of applications
in many fields ranging from computer science to biology and
sociology \cite{DMW2003,OC2004,XCHH2004,XJC2004,BA99}. In this
respect, three important results have been crystallized: First, it
has been found that the degree distribution contains important
information about the nature of the network, for many large
networks following the exponential distribution and the power-law
distributions. Second, most networks have the so-called small
world property \cite{DMbook}, which means that the average
distance between different pairs of nodes is rather small. Third,
real networks display a degree of clustering coefficient higher
than expected for random networks \cite{DMbook,OC2004}. Finally,
the assortative mixing is studied to answer why social networks
are different from other types of networks \cite{NP2003}.

The scientific studies can be considered as being organized within a
network structure, which has a significant influence on the observed
study collective behaviors. The viewpoints of complex networks are
of interest in studying scientific study networks, to uncover the
structural characteristics of the networks built on RAN. In this
paper, the research areas of natural science basic research is
studied from the point view of complex network \cite{F}. In the fund
management department, such as National Natural Science Foundation
of China (NSFC), the research areas are denoted by the code system,
which have the tree structure to demonstrate the inclusion relation
between the research areas, such as Physics--$>$statistical
physics--$>$complex network. The leave codes of the code system
always represent the research areas more specially. To make the
network reflect the reality more accurately, in this paper, the
nodes are defined as the leave nodes of the code system. Because the
scientists can fill in the fund proposal two codes: the first
application code and the second one, then if one requisition paper
filled in two different codes one can consider that the research
work is cross the two research areas. The proposals filled in only
one code are not considered in RAN. By this definition, there are
371, 349, 367, 400, 456, 544 nodes in RAN from 1999 to 2004. Three
complementary approaches allow us to obtain a detailed
characterization. First, empirical measurements allow us to uncover
the topological measures that characterize the network at a given
moment, as well as the time evolution of these quantities.
%The results indicate that the network is SED affected by both internal
%and external links.
Second, the average distance of RAN decreases
with time, which means that the distance between any pairs of
research areas is getting short. Third, the average clustering
coefficient increases with time, which means that the neighbors of
one research area would like to be connected with each other.

%It should be noted that the topology of RAN does not have
%significant change in a long period, because NSFC would not like to
%adjust its research areas frequently. Therefore, to the specific
%interest of this paper, we attempt to uncover the topological
%properties of RAN, leaving a further study of its complex
%dynamical features to the near future.

This paper is organized as follows: In section 2, the topological
characteristics of RAN, such as the degree distribution, clustering
coefficient, average path length, assortative coefficient and the
relationship between the clustering coefficient and the degree $k$
are investigated and visualized. In section 3, the conclusion and
discussion are given.

\section{Data Analysis of RAN}
In this section, the topology and dynamics of the empirical network
are investigated. The parameters that are crucial to the
understanding of the topology of RAN are extracted. The analysis of
the data could provide the development trend of Chinese natural
scientific basic research system.
\begin{figure}%[ht]
    \begin{center}
       {\includegraphics[width=0.6\textwidth]{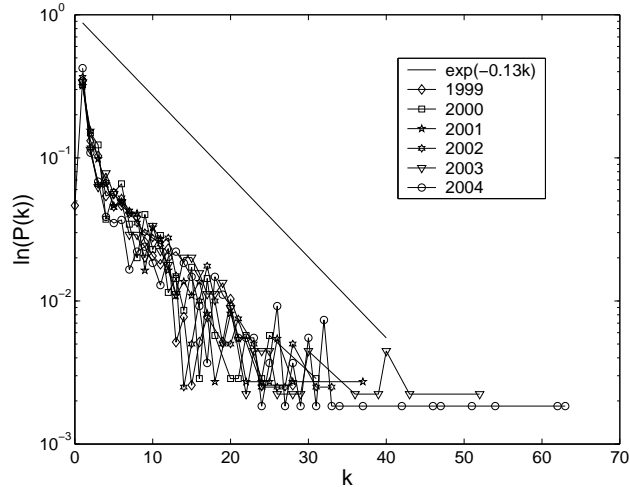}}
 \caption{The degree distribution of RAN from 1999 to 2004.}
 \end{center}
\end{figure}

\subsection{Degree distribution follows the exponential form}
The degree distribution $P(k)$ presents the probability that a
randomly selected node has $k$ links, which has been studied
extensively for various networks. Networks for which $P(k)$ has a
power-law tail, are known as scale-free networks
\cite{BA99,BAJ99}. On the other hand, classical network models,
including the Erd$\check{o}$s-R$\acute{e}$nyi model
\cite{ER59,ER60} and the Watts and Strogatz model \cite{WS98} have
an exponentially decaying $P(k)$ and are collectively known as
exponential networks. Another observed degree distribution form of
real-life networks\cite{XCHH2004,Collaboration}, named Stretched
exponential distribution (SED) \cite{LS1998} is of the form
$P(x)dx=\mu(x^{(\mu-1)}/x_0^{\mu}){\rm exp}(-(x/x_0)^{\mu})dx$ and
its accumulative distribution is $P(x)={\rm exp}(-(x/x_0)^{\mu})$,
which can be stated as ${\rm ln}P(x)\sim x^{\mu}$. Obviously, SED
degenerates to an exponential distribution when $\mu$ is close to
1 and to a power law when $\mu$ is close to 0. When $\mu$ is
between 0 and 1, the degree distribution is between a power law
and an exponential function. The data on a single-logarithmic
plane show that the degree distribution of RAN can well fitted of
an exponential form, see Fig. 1 and Fig. 2.

%The degree distribution of the data indicates that RAN is SED, see
%Fig. 1 and Fig. 2.
%There is an elbow  value
%around $k_c=10$ for the log-log plot of $P(k)$ of the six years. As
%there is a lot of scatter, we also determine the corresponding
%cumulative distribution $P_c(k)$. The latter distribution is fitted
%by two different exponentials, which is presented in Table 1.

%\begin{center}
% { Table 1 The exponent of the cumulative distribution $p_c(k)$}\\
% \begin{tabular}{c c c c c c c }\hline
%   $k_c$  & 1999    & 2000   & 2001   & 2002   & 2003   & 2004   \\ \hline
% $k_c>10$& -3.6574 &-3.4737 & -3.2438&-3.3620 &-2.7161 & -2.4889\\ %\hline
% $k_c<10$& -0.8996 &-0.8045 & -0.7709&-0.7094 &-0.6428 & -0.6093\\  \hline
% \end{tabular}
%\end{center}
It seems difficult to find a common function to fit the empirical
network because of the fluctuations. However, the accumulative
distribution of RAN has less fluctuations and is more stable. The
distribution of all the data from 1999 to 2004 can be fitted by an
exponential distribution, given by $P(k)={\rm exp}(0.13k)$. The hub
nodes of RAN and their degrees from 1999 to 2004 are demonstrated in
Table 1.

\begin{figure}%[ht]
    \begin{center}
       {\includegraphics[width=0.8\textwidth]{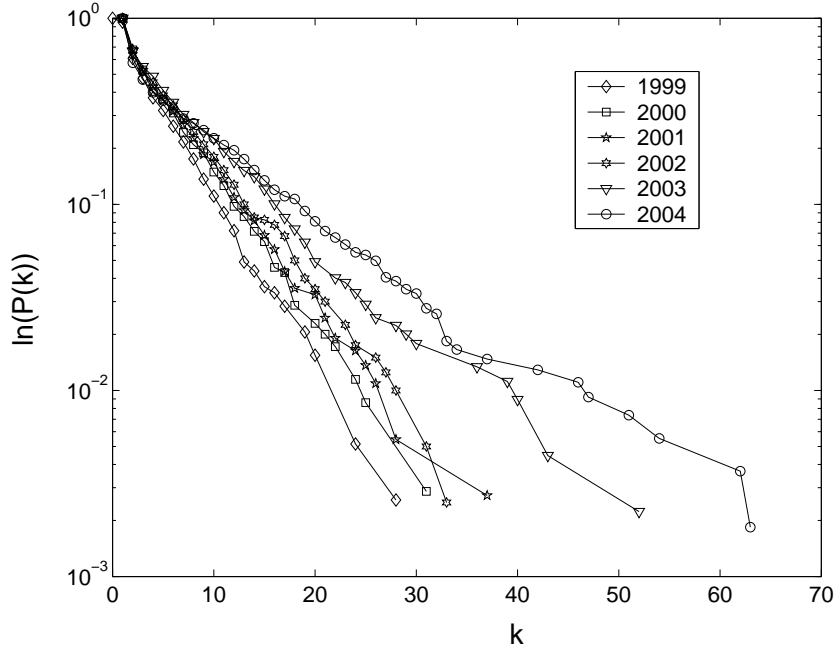}}
 \caption{The cumulation degree distribution of RAN from 1999 to 2004.}
 \end{center}
\end{figure}

\begin{figure}%[ht]
    \begin{center}
       {\includegraphics[width=0.8\textwidth]{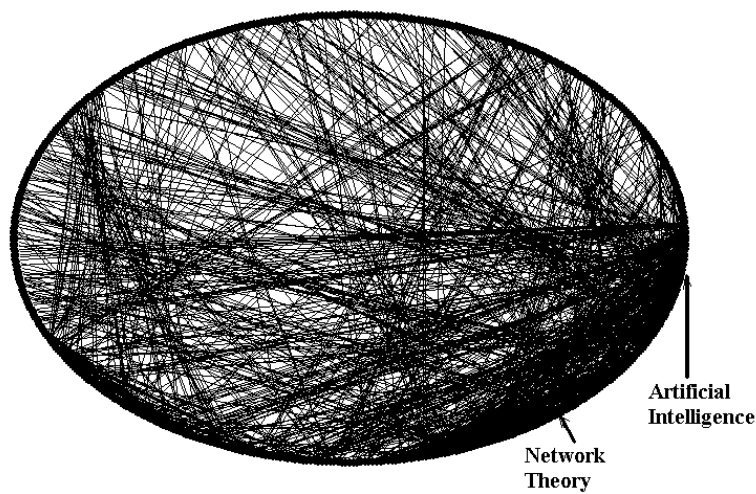}}
 \caption{The network diagram of 2004. The marked points are
the research areas of Network theory and Artificial Intelligence.}
 \end{center}
\end{figure}

%The cumulative distribution implies that the original distribution
%has also two exponential regimes \cite{New}. The existence of the
%threshold $k_c$ indicates that most parts of the research areas have
%connectivity less than $k_c$ and a few ones have connectivity more
%than $k_c$. The research areas, which have less than $k_c$ degree,
%probably lay emphasis on the theoretical study, so they only connect
%to limited correlational research areas. The others, which have lots
%of connectivity, probably lay emphasis on the application or
%probably are new risen research areas. Therefore, they would like to
%cross many basic and applicative research areas, such as the
%intelligence artificial and the graphic processing. The network
%diagram of 2004 is demonstrated in Fig. 3.

\begin{center}
{Table 1 The hub nodes of RAN and their degrees from 1999 to
2004.}
  \begin{tabular}{c c c}\hline
             &  Hub nodes  & degree \\ \hline
   1999      &  Computer-aided design & 28 \\ %\hline
   2000      & Intelligent information processing(IIP) & 31 \\ %\hline
   2001      & IIP & 37  \\
   2002      & Pattern cognition, Artificial intelligence(PC \& AI) & 33 \\
   2003      &PC \& AI &  52 \\
   2004      & PC \& AI & 63\\ \hline
 \end{tabular}
\end{center}

\subsection{Average distance decreases with time}
The ability of two nodes, $i$ and $j$, to communicate with each
other depends on the length of the shortest path $d_{ij}$ between
them. The average of $d_{ij}$ over all pairs named average distance
$D=\frac{1}{N(N-1)}\sum_{ij}d_{ij}$.

\begin{figure}[ht]
    \begin{center}
       {\includegraphics[width=0.6\textwidth]{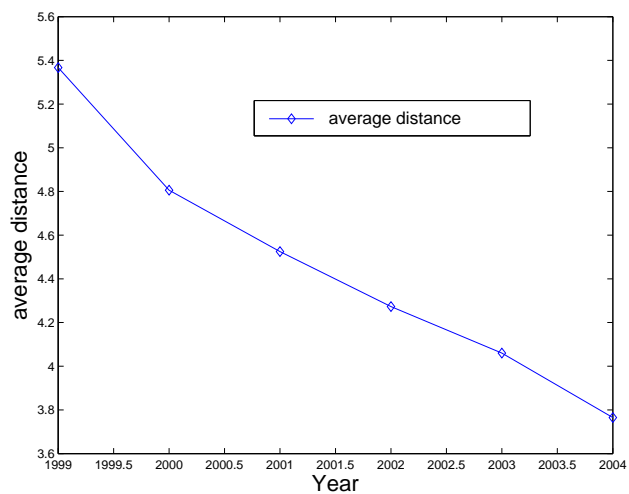}}
 \caption{Average distance of RAN.
 The separation is computed on the data up to the indicated year.
 }
 \end{center}
\end{figure}
Fig. 4 shows that $D$ decreases with time, which is highly
surprising because most network models so far predict that the
average distance should increases with system size. The decreasing
trend observed by us could have different origins. It is possible
that the increasing interconnectivity decreases the average
distance. In analogy with the social networks, one can say that
there are 3.8 degrees of separation between the research areas.

One can note the continuous declining of the average distance and
the more connected nature of RAN fields expressed by a smaller
separation. The monotone declining indicates that
%even longer time
%interval is needed to reach the asymptotic limit, in which different
%relevant quantities take up a stationary value. This finding
%provides the theoretical evidence that
the scientists tend to study in crossing areas.
%and the research areas tend to be integrated.

\subsection{Average Clustering coefficient increases with time}
\begin{figure}[ht]
    \begin{center}
       {\includegraphics[width=0.6\textwidth]{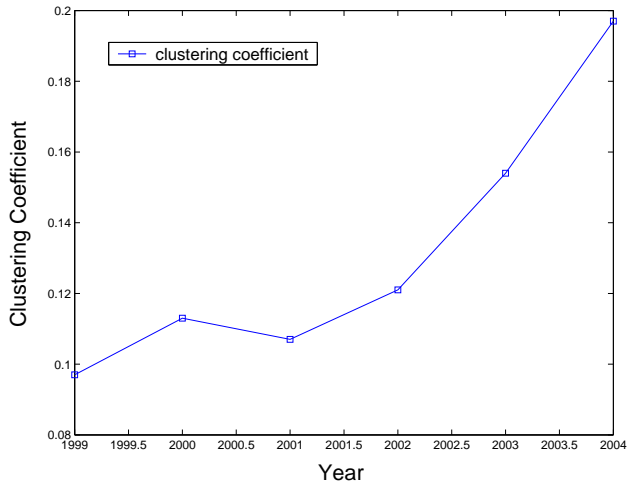}}
 \caption{Average clustering coefficient of RAN, determined for the data up to
 the indicated year.}\label{Fig4}
 \end{center}
\end{figure}

The clustering coefficient $C_i$ can be defined as follows
\cite{WS98}: pick a node $i$ that has links to $k_i$ nodes in the
system. If these $k_i$ nodes form a fully connected clique, there
are $k_i(k_i-1)/2$ links between them, but in reality we find much
fewer. Denote the number of links that connect the selected $k_i$
nodes to each other by $E_i$. The clustering coefficient for node
$i$ is then $C_i=2E_i/[k_i(k_i-1)]$. The average clustering
coefficient for the whole network is obtained by averaging $C_i$
over all nodes in the system $C=\frac{1}{N}\sum_{i=1}^NC_i$. In
simple terms, the clustering coefficient of RAN indicates the
probability that two different nodes connect to the same node. The
data of RAN indicates average clustering coefficient decrease with
time, which can be seen from Fig. \ref{Fig4}.

%The two different nodes which connect to the same node

\subsection{Average degree increases}

The number of nodes of RAN increases with time because of the
arrival of new research areas. The growth of total number of links
comes from two parts. One is the connections generated by new
research areas with old ones, the other is the new connections
between old nodes. A quantity characterizing the network's
interconnections is the average degree $\langle k \rangle$, denoting
the average number of links per node. The time dependence on
$\langle k \rangle$ of RAN is shown in Fig. 6.

\begin{figure}[ht]
    \begin{center}
       {\includegraphics[width=0.6\textwidth]{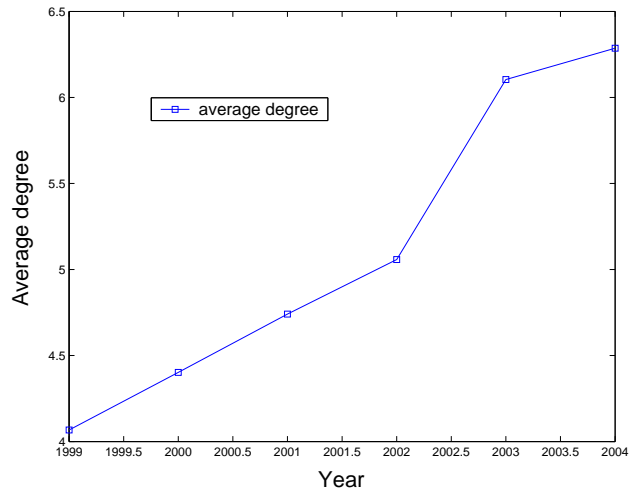}}
 \caption{Average number of links per node for RAN. Results are computed
 on the data up to the given year.}
 \end{center}
\end{figure}

\subsection{Assortative Coefficient}
The assortative coefficient $A$ measures the tendency of a network
to connect vertices with the same or different degrees
\cite{N2002,NP2003}. If $A>0$, the network is said to be
assortative, non-assortative when $A=0$ and disassortative when
$A<0$. $A$ is defined as follows
\begin{equation}\label{2.51}
A=\frac{1}{\sigma^2_q}\sum_j \sum_k (e_{jk}-q_jq_k),
\end{equation}
\begin{center}
 { Table 2 \quad Temporal evolution of some quantities of RAN.
  In the first column, $v$ is the number of vertices,
 $\overline{k}$ is the mean connectivity, $C$ is the clustering coefficient,
 and $D$ is the average shortest path length}\\
 \begin{tabular}{c c c c c c c c c} \hline\hline
      & 1999 & 2000 & 2001 & 2002 & 2003 & 2004 \\ \hline
   $v$& 371  & 349  & 367  &  400 &  456 & 544  \\
 %  $e$& 388  &      &      &      &      &      \\
$\overline{k}$& 4.264&4.768&4.986 &5.235 &6.197& 6.379      \\
   $C$&0.097 &0.113 &0.107 &0.121 &0.154 &0.197 \\
   $D$&5.367 &4.806 &4.525 &4.273 &4.060 &3.765  \\ %\hline
   ${\rm isolate}\ \ {\rm node}$& 18 & 29 & 19 & 14 & 7 & 8 \\
   $A$&-0.0187  &0.0033 &0.0796 &-0.0084&-9.7050e-004&-0.0899  \\ \hline\hline

 \end{tabular}
\end{center}
where $e_{jk}$ is the probability that a randomly chosen edge has
vertices with degree $j$ and $k$ at either end, $q_k=\sum_j e_{jk}$
and $\sigma^2_q=\sum_k k^2q_k-(\sum_k kq_k)^2$. The possible values
of $A$ lies in the interval $-1\leq A\leq 1$. Fig. 7 shows that the
assortative coefficient begin to decline from 2001 and it decreases
to the negative value from 2002. This indicates that RAN tend to be
disassortative network.

\begin{figure}[ht]
    \begin{center}
       {\includegraphics[width=0.7\textwidth]{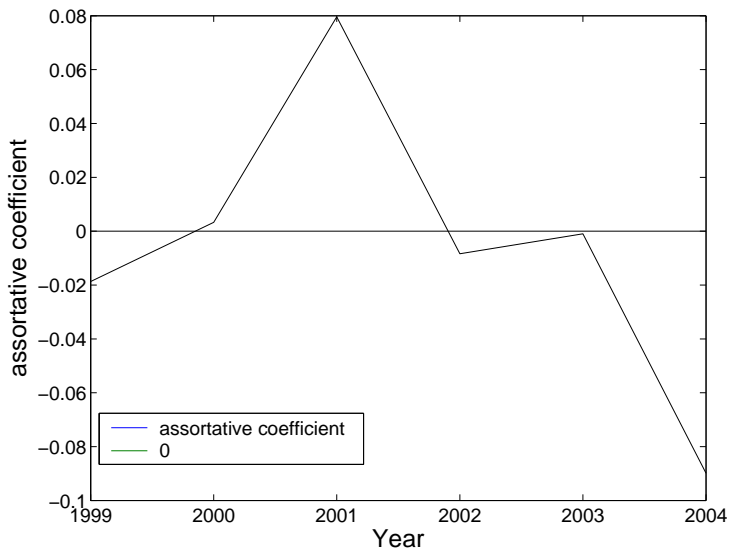}}
 \caption{Assortative coefficient of RAN. Results are computed
 on the data up to the given year.}
 \end{center}
\end{figure}

\subsection{Dependence between clustering coefficient and degree}
This intrinsic hierarchy can be characterized in a quantitative
manner using the recent findings of Dorogovtsev {\it et. al}
\cite{DGM} and Ravasz {\it et. al} \cite{RB}. The hierarchical
organization, implying that small groups of nodes organize in a
hierarchical manner into increasingly large groups. In RAN, the
relationships between clustering coefficient and the degree of
clustering characterizing from 1999 to 2004 are demonstrated in Fig.
8.

\begin{figure}[ht]
    \begin{center}
       {\includegraphics[width=0.7\textwidth]{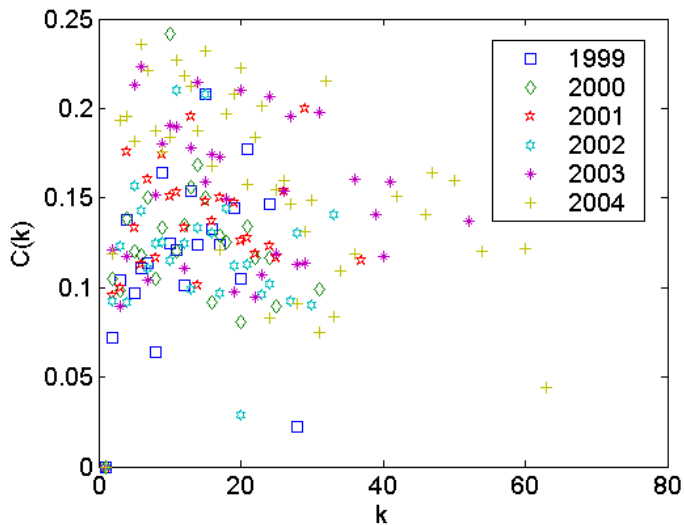}}
 \caption{Dependence between clustering coefficient and $k$ from 1999 to 2004.}
 \end{center}
\end{figure}
From Fig. 8, one can see that there are no dependence between
clustering coefficient and $k$ from 1999 to 2004, which means that
there is no hierarchical organization in RAN.
%But the data of 2004
%demonstrates that the clustering coefficient of a node with $k$
%links follows the scaling law
%\begin{equation}
%C(k)\sim k^{-1}.
%\end{equation}
%This scaling law quantifies the coexistence of a hierarchy of nodes
%with different degrees of clustering. Those at the center of one
%node have $k=63$ and $C(63)=0.04331$, while those at the center of
%the 7-node modules have $k=18$ and $C(18)=0.1972$, indicating that
%the higher a node¡¯s degree, the smaller is its clustering
%coefficient, asymptotically following the reciprocal law.

\section{Conclusion and Discussion}
In this paper, we have constructed a research areas network and
pointed out that such a network falls into the exponential
distribution and small-world networks categories. With the
observation on the section 2, RAN tend to be a disassortative
network. The average distance of RAN decreases with time, while
the clustering coefficient increases with time. This indicates
that the research areas are tend to be crossed studied by the
scientists. The dependence between clustering coefficient and $k$
indicates that there is no hierarchical organization in RAN from
1999 to 2004.

The characteristics of RAN may lies in the following reasons.
\begin{description}
\item[1.] The code system is changeless in a period of time, but the
scientific study develops very fast, there are new research subjects
emerge everyday. So some scientists can not find the suitable code
which could describe their research content but only fill in the
proposal two corelational codes.

\item[2.] The research subjects become more complexity, which can not
be solved by one or two fields of knowledge. So the contents actuate
the scientists who studied them to fill in their proposal two codes.
\end{description}

%Since the topology structure of RAN is not determined by any
%individual,

% In RAN, the small groups
%of nodes tend to organize in a hierarchical manner into increasingly
%large groups.

We hope that the work presented here may stimulate further research
on this subject. Some opened questions are, for instance, whether
the results obtained by RAN held for different countries?
%or, perhaps, for different departments?
What is the evolving mechanism of RAN?

%It should be noted again that the topology of RAN does not have
%significant change in a long period.

%\textbf{\{Because the scale-free featured web of the research
%areas is almost fixed, the topological from RAN on the currency
%crises and research areas of the scientists studied become more
%attention global.\}}

\subsection*{Acknowledgement}
The authors are grateful to Dr. Qiang Guo, Bing Wang, Zhong-Zhi
Zhang and Tao Zhou for their valuable comments and suggestions,
which have led to a better presentation of this paper. This work
has been supported by the National Science Foundation of China
under Grant Nos. 70431001 and 70271046.


\begin{thebibliography}{0}
\bibitem{F}%30
Data were gathered from the fund management deparment of China.
\end{thebibliography}
\end{document}